\documentclass{elsart3p}

\usepackage{natbib}

\usepackage{graphicx}
\usepackage{amssymb}

\begin{document}

\begin{frontmatter}

% Title, authors and addresses

% use the thanksref command within \title, \author or \address for footnotes;
% use the corauthref command within \author for corresponding author footnotes;
% use the ead command for the email address,
% and the form \ead[url] for the home page:
% \title{Title\thanksref{label1}}
% \thanks[label1]{}
% \author{Name\corauthref{cor1}\thanksref{label2}}
% \ead{email address}
% \ead[url]{home page}
% \thanks[label2]{}
% \corauth[cor1]{}
% \address{Address\thanksref{label3}}
% \thanks[label3]{}

\title{Is irradiation important for the secular evolution of
       low-mass X-ray binaries?}

% use optional labels to link authors explicitly to addresses:
% \author[label1,label2]{}
% \address[label1]{}
% \address[label2]{}

\author{H. Ritter}

\address{Max-Planck-Institut f\"ur Astrophysik, 
         Karl-Schwarzschild-Str. 1, 
         D-85741 Garching, 
         Germany}

\begin{abstract}
It is argued that irradiation in low-mass X-ray binaries (LMXBs)
caused by accretion-generated X-rays can not only change the optical
appearance of LMXBs but also their outburst properties and
possibly also their long-term evolution. Irradiation during an
outburst of the outer parts of the accretion disc in a transient LMXB
leads to drastic changes in the outburst properties. As far as the
secular evolution of such systems is concerned, these changes can
result in enhanced loss of mass and angular momentum from the system
and, most important, in neutron star LMXBs in a much less efficient
use of the transferred matter to spin up the neutron star to a
ms-pulsar. Irradiation of the donor star can destabilize mass transfer
and lead to irradiation-driven mass transfer cycles, i.e. to a
secular evolution which differs drastically from an evolution in which
irradiation is ignored. It is argued that irradiation-driven mass
transfer cycles cannot occur in systems which are transient because of
disc instabilities, i.e. in particular in long-period LMXBs with a
giant donor. It is furthermore shown that for irradiating either the
disc or the donor star, direct irradiation alone is insufficient.
Rather, indirect irradiation via scattered accretion luminosity must
play an important role in transient LMXBs and is, in fact, necessary
to destabilize mass transfer in short-period systems by irradiating
the donor star. Whether and to what extent irradiation in LMXBs does
change their secular evolution depends on a number of unsolved
problems which are briefly discussed at the end of this article.  
\end{abstract}


\end{frontmatter}

% main text
\section{Introduction}
\label{Intr}

Accretion onto either the neutron star (NS)or the black hole (BH) 
component of a low-mass X-ray binary (LMXB) liberates copious amounts
of energy, mainly as X-rays which, in turn, can irradiate either the
accretion disc or the donor star of the binary system. Only gradually
was it realized that irradiation in a LMXB is not just a side effect
which changes its appearance but, more important, that it also changes
the conditions for the occurrence of outbursts in X-ray transients
\citep{van_Paradijs94} and their characteristics 
\citep{King&Ritter98}, and possibly the long-term evolution of these
objects as well \citep{Podsiadlowski91}. Given ample evidence that
irradiation does influence LMXBs in different ways the question
arises to what extent it is of importance for their long-term
evolution.  

The consequences would be drastic indeed if the situation modelled by 
\citet{Podsiadlowski91}, namely spherically symmetric irradiation of
the donor star, would really apply. However, as has been argued at
length by \citet{RZK00} (hereafter RZK), this is probably not the case.
Rather the synchronously rotating donor star intercepts accretion
luminosity essentially only on the hemisphere facing the source while
the remaining parts are in the shadow and undisturbed. Whereas 
irradiation is of little consequence for ``hot'' stars (with an
effective temperature $T_{\rm eff} \gtrsim 6500$~K) having a radiative
envelope, this is not the case for ``cool'' stars ($T_{\rm eff}
\lesssim 6500$K) with a convective envelope. As has been shown by RZK
the consequences of one-sided irradiation of a cool star are rather
subtle and much less drastic than what \citet{Podsiadlowski91} found
for the case of spherically symmetric irradiation. In addition, it was
found that for binary parameters which are typical for at least a
significant fraction of the observed LMXB ensemble irradiation 
of the donor star by accretion luminosity could possibly force mass
transfer to undergo what we will refer to as irradiation-driven mass
transfer cycles. In this case, phases of high mass transfer, driven by
the thermal expansion of the convective envelope of the irradiated
donor, alternate with phases with low or no mass transfer during which
the donor readjusts towards thermal equilibrium of the unirradiated
star. Details about irradiation-driven mass transfer cycles may be
found in RZK, \citet{B&R04}, and references therein. 

As we have seen there are basically two ways in which
accretion-generated irradiation can interfere with the evolution of a
LMXB: One way is by irradiation of the accretion disk which, in turn,
changes the conditions for the occurrence of disc instabilities and
the outburst properties in transient systems (for a review see e.g. 
\citet{Lasota01}). Although disc instabilities occur on timescales
much shorter than the evolutionary timescale of a LMXB, they can
nevertheless change the outcome of binary evolution, as we shall see. 
The other way in which irradiation can affect binary evolution is by
forcing it to undergo irradiation-driven mass transfer cycles. 

In the following I shall examine in more detail the conditions for the
occurrence of these two processes and their consequences for the
long-term evolution of LMXBs.  

\section{Sources and receptors of radiation in a LMXB}
\label{Sources}

Before going to examine the processes mentioned above it is perhaps
useful to have a look at the possible sources and receptors of
radiation in a LMXB and the conditions that have to be fulfilled in
order for irradiation to play a role at all. 

The possible sources and receptors of radiation can be conveniently
subdivided into: (1) the compact star, (2) the inner parts of the
accretion disc and the accretion disc corona, (3) the outer parts of
the disc, and finally (4) the donor star. (The ``inner parts of the
disc'' include a possible boundary layer and those parts of the disc
where most of the binding energy is liberated.) Of course, not all
combinations of source $(i),~i=1, \dots,4$ irradiating receptor 
$(j),~j=1, \dots,4$ are possible or of relevance. In order to be of
importance the radiation produced by the source has to meet two
conditions: first it has to be hard (penetrating) enough to
reach the receptor and penetrate below its photosphere, and second its
flux density at the receptor should be at least comparable to if not 
exceeding the receptor's intrinsic flux. Excluding self-irradiation
this leaves us essentially with the following combinations:
Irradiation of the inner and outer disc as well as the donor star by
the compact star (here the neutron star), and irradiation of the outer
disc and the donor star by the inner disc and corona (which is of
relevance if the compact object is a black hole). 

\section{Effects of irradiation} 
\label{Effects} 
\subsection{Effects on the disc}
\label{DiscEffects} 
The main effects of irradiating the accretion disc which are of
relevance here can be summarized as follows (for details see e.g.
\citet{van_Paradijs94}, \citet{King&Ritter98}, \citet{Lasota01}, 
\citet{Ritter&King01}): 
\begin{itemize} 
\item Whereas viscous heating keeps the effective temperature 
      $T_{\rm eff}$ of the disc above the hydrogen ionization
      temperature $T_{\rm H} \approx 6500$~K inside a radius 
\begin{equation}
R_{\rm h, visc} \approx 0.66~{\rm R_{\odot}}~{m_{\rm c}}^{1/3} 
                    {\dot M_{\rm tr, -8}}^{1/3}\,,
\label{Rhvisc}
\end{equation}
      where $m_{\rm c}$ is the mass of the compact star in 
      ${\rm M_{\odot}}$ and $\dot M_{\rm tr, -8}$ the mass transfer
      rate in units of $10^{-8} {\rm M_{\odot}/yr}$, irradiation,
      following \citet{King&Ritter98}, keeps $T_{\rm eff}$ in the disc
      of a neutron star LMXB above $T_{\rm H}$ inside a radius 
\begin{equation}
R_{\rm h, irr} \approx 7.1~{\rm R_{\odot}}~{\dot M_{\rm tr, -8}}^{1/2}\,.
\label{RhirrNS}
\end{equation}
      As a consequence of this, for a binary of given orbital period
      $P_{\rm orb}$, stationary accretion is possible for lower mass 
      transfer rates than without irradiation, or for longer
      $P_{\rm orb}$ for a given $\dot M_{\rm tr}$. 

      For a black hole LMXB, again using parameters of
      \citet{King&Ritter98}, one obtains 
\begin{equation}
R_{\rm h, irr} \approx 2.7~{\rm R_{\odot}}~{\dot M_{\rm tr, -8}}^{1/2}\,.
\label{RhirrBH}
\end{equation}
\item In cases where disc accretion is non-stationary, e.g. if the
      donor is a giant and $P_{\rm orb}$ long, in an outburst
      irradiation delays the return to quiescence. As a consequence, 
      the disc is emptied to a much larger degree and this, in
      turn, results in a much longer quiescence (mass accumulation
      phase) beteween outbursts. 
\item Furthermore, in long-period systems a much larger part of the
      disc, i.e. much more mass is actively involved in an outburst.
      This results in  much higher mass flow rates $\dot M_{\rm d}$
      through the disc, rates which can be much larger than the
      Eddington accretion rate $\dot M_{\rm Edd}$ of the compact star.
      And this, in turn, leads to significant loss of mass and angular
      momentum from the binary system. 
\end{itemize}

\subsection{Effects on the donor star}
\label{DonorEffects}

\begin{figure}
\begin{center}
\includegraphics*[width=7.5cm]{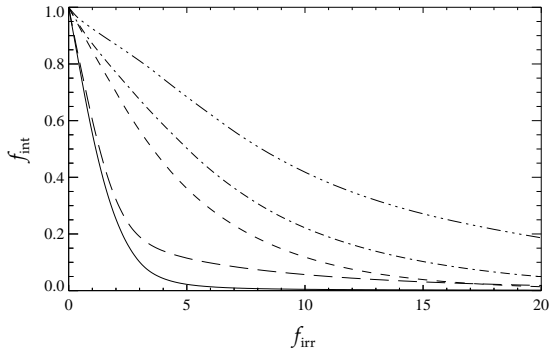}
\end{center}
\caption{$f_{\rm int} = F_{\rm int}/F_0$ ($F_0$ being the intrinsic
flux of the unirradiated star) as a function of $f_{\rm irr}
= F_{\rm irr}/F_0$ for five different stellar models characterized by
their evolutionary state and the following combination of mass and
central hydrogen abundance $(M, X_{\rm c})$: full line: ZAMS, (0.3,
0.71); long-dashed line: ZAMS, (0.5, 0.71); short-dashed line: ZAMS,
(0.8, 0.71); dash-dotted line : near the TAMS, (0.45, 0.05), and
dot-dash-dotted line: giant with a radius of $R=25.8 {\rm R_{\odot}}$,
(0.8, 0.0). From \citet{B&R04}.}
\label{Fint(Firr)}
\end{figure}

\begin{figure}
\begin{center}
\includegraphics*[width=8.5cm]{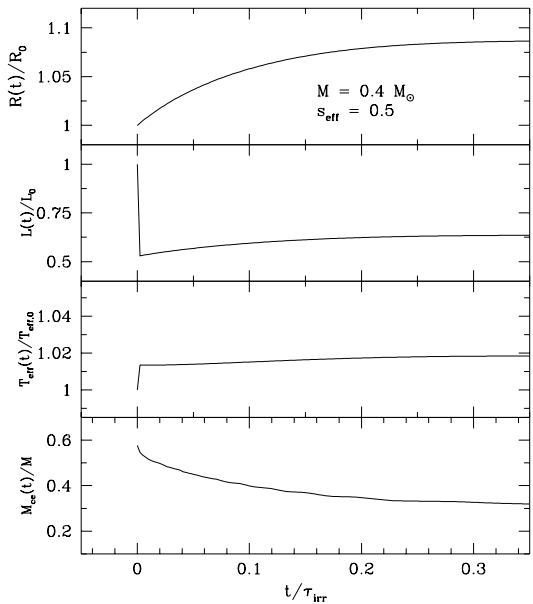}
\end{center}
\caption{Thermal relaxation of a $0.4 {\rm M_{\odot}}$ main sequence
star after blocking the energy loss over a fraction $s_{\rm eff}=0.5$
of its surface at time $t=0$. Top frame: radius $R$, second frame:
luminosity $L$, third frame: effective temperature $T_{\rm eff}$ of
the unirradiated part, and bottom frame: the relative mass
$M_{\rm ce}/M$ of the convective envelope as a function of time. Time
is measured in units of the timescale on which the radius grows at
$t=0$. $R_0$, $L_0$, and $T_{\rm eff,0}$ are respectively the values
of $R$, $L$ and $T_{\rm eff}$ immediately before the onset of the
blocking of energy outflow (from RZK).}
\label{th.relax}
\end{figure}

The most important effects resulting from irradiating the donor star
can be summarized as follows (for details see
e.g. \citet{Podsiadlowski91}, RZK and \citet{B&R04}): 
\begin{itemize}
\item For ``hot''stars ($T_{\rm eff} \gtrsim 6500$~K, radiative
      envelope) irradiation is of little consequence. 
\item For cool stars with a convective envelope ($T_{\rm eff} \lesssim
      6500$~K) irradiation, by lowering the superadiabatic temperature
      gradient in the subphotospheric layers, hinders the star's
      energy loss from the interior through the irradiated parts of
      its surface. As a result, the flux $F_{\rm int}$ which such a
      star can lose from its interior is the lower the higher the
      irradiating flux $F_{\rm irr}$\footnote{To be precise: 
      $F_{\rm irr}$ hereafter denotes the perpendicular component of
      that fraction of the external irradiating flux which is absorbed 
      below the photosphere.}. As an example, the relation 
      $F_{\rm int}(F_{\rm irr})$ as derived from numerical
      calculations \citep{Hameury&Ritter97} is shown in
      Fig.~\ref{Fint(Firr)} for a number of cool low-mass stars.
\item Upon onset of irradiation, an initially undisturbed cool star
      will expand on the timescale $\tau \approx \tau_{\rm ce}/
      {s_{\rm eff}}$,  where $\tau_{\rm ce}$ is the thermal timescale
      of the convective envelope, and $s_{\rm eff}$ the effective
      fraction of the stellar surface through which energy loss from
      its interior is blocked. An example of a numerical calculation
      of the thermal relaxation process of a low-mass ($0.4
      {\rm M_{\odot}}$) star is shown in Fig.~\ref{th.relax}. In a
      semi-detached binary this expansion drives additional mass
      transfer and thus leads to increased accretion luminosity and
      hence irradiating flux. 
\item Because the thermal relaxation process, i.e. the radius increase,
      saturates with time (see top panel in Fig.~\ref{th.relax}) and in 
      amplitude (meaning that no more than the total intinsic flux can
      be blocked), in a semi-detached binary this leads to the
      possibility of irradiation-driven mass transfer cycles mentioned
      earlier. Whether such mass transfer cycles do appear depends on
      whether mass transfer is thermally stable. A detailed discussion
      of this problem is beyond the scope of this paper. For this the
      reader is referred to \citet{KFKR96}, \citet{KFKR97}, RZK, and 
      \citet{B&R04}. If mass transfer is unstable and cycles do occur,
      then phases of high mass transfer alternate with phases of low
      or no mass transfer during which the donor readjusts towards
      thermal equilibrium of the unirradiated star. 
\item For a partially irradiated cool star the relative radius
      increase saturates at a value 
\begin{equation}
      {\Delta R}/R \approx (1 - s_{\rm eff})^{-\rho} - 1\,, 
\label{DR/R}
\end{equation} 
      where $\rho \approx 0.1$ for low-mass main sequence stars, and 
      $\rho \approx 0.5$ for giants. Thus, for any value of  
      $s_{\rm eff}$ which is not very close to unity the relative
      radius increase is much smaller than what 
      \citet{Podsiadlowski91} had found for saturated, spherically
      symmetric irradiation, whereby the stellar envelope becomes
      radiative.  
\item Upon sustained or slowly varying irradiation an isolated
      irradiated star attains a new thermal equilibrium radius 
\begin{equation}
      R_{\rm e}(s_{\rm eff}) \approx R_{\rm e}(0)~ 
                                     (1 - s_{\rm eff})^{-\rho}.  
\label{Re(s)}
\end{equation}
      In a mass transferring binary the irradiated donor cannot  
      attain thermal equilibrium. Nevertheless, for a given mass 
      and mass loss rate the irradiated star is systematically
      oversized compared to the case where irradiation is ignored.
\end{itemize}

\section{Consequences for the secular evolution of neutron star LMXBs}
\label{NSLMXBs}
\subsection{Consequences from disc irradiation}
\label{NSLMXBs.1}
Depending on whether the disc radius $R_{\rm d} \lessgtr 
R_{\rm h}(\dot M_{\rm tr})$ disc accretion is stable ($<$) or
unstable ($>$). Since $R_{\rm h, irr} > R_{\rm h, visc}$ for all values
of $\dot M_{\rm tr}$ of interest (cf. Eqs. (1) and (2)), disc
irradiation widens the parameter space for systems with stable disc
accretion. Using standard parameters for describing an irradiated disc
\citep{King&Ritter98} and assuming $R_{\rm d}$ to be 80\% of the
Roche radius of the compact component, the critical orbital period, 
$P_{\rm crit}$, for which $R_{\rm d} = R_{\rm h, irr}$, is 
\begin{equation} 
P_{\rm crit} \approx 9.8{\rm ~d}~ {m_{\rm c}}^{-0.675} {m_{\rm d}}^{0.175} 
                     {\dot M_{\rm tr, -8}}^{3/4}\,,
\label{PcritNS}
\end{equation}
where $m_{\rm d}$ is the mass of the donor in ${\rm M_{\odot}}$.
Accordingly, most of the short-period neutron star LMXBs ($P_{\rm orb}
\lesssim 2$~d) should have stationary discs. And for such systems it
appears, at least at first glance, that irradiation is of little
consequence for their long-term evolution. However, as we shall see
below, it is exactly these systems for which irradiation-driven mass
transfer cycles are most likely to occur, and with them the
consequential changes of their long-term evolution. 

If, on the other hand, $P_{\rm orb} > P_{\rm crit}$ disc accretion is
unstable and such systems are X-ray transients. As we have already
noted above, because $R_{\rm h, irr}$ is significantly bigger than 
$R_{\rm h, visc}$ (cf. Eqs. (1) and (2)), a much larger part of the
disc is involved in an outburst than would be the case without
irradiation. Therefore, much more mass is actively involved in an
outburst. Because of irradiation, it also takes longer for the disc to
return to quiescence. This happens only after the disc is almost
totally emptied. For a given mass transfer rate this means that the
quiescene, i.e. the mass accumulation time between two consecutive
outbursts, is correspondingly longer. In addition, the mass flow rate
through the disc during an outburst is the larger the longer the
orbital period. In fact, for transient systems the mass flow rates are
typically much above the Eddington rate of a neutron star (see e.g. 
\citet{Ritter&King01}). In this context it is important to note that
in long-period systems, i.e. with $P_{\rm orb} \gtrsim 20$~d,
undergoing nuclear timescale-driven mass transfer from a giant, even
the time-avaraged mass transfer rate which is roughly $\propto  
P_{\rm orb}$ exceeds the Eddington accretion rate of a neutron star
(e.g. \citet{Ritter99}). This, in turn, has two consequences for the
evolution of a LMXB: First, super-Eddington mass flow rates lead to
substantial loss of mass and angular momentum from the binary system
and thus to an evolution which differs from that of an LMXB where
irradiation has been ignored. Second, super-Eddington mass flow rates
also mean that only a (small) fraction of the transferred matter can
be accreted by the neutron star. And that fraction is the smaller the
longer $P_{\rm orb}$. In this way irradiation of the disc hinders or,
in long-period systems ($P_{\rm orb} \gtrsim {\rm few}~10^2~{\rm d})$,  
may even prevent the spin-up of the neutron star and thus the
formation of millisecond pulsars \citep{Ritter&King01}. To make
matters worse, at least as far as the formation of millisecond pulsars
is concerned, during the long-lasting quiescence the neutron star
could be significantly spun down by becoming a propeller (see below
Sect. \ref{propeller}). 

\subsection{Consequences from irradiating the donor star} 
\label{NSLMXBs.2}
As has already been detailed in Sect. \ref{DonorEffects} irradiating a
low-mass donor star in a LMXB can destabilize mass transfer and give
rise to irradiation-driven mass transfer cycles. If mass transfer is
thermally stable and no mass transfer cycles do occur, irradiation is
of little consequence for the long-term evolution of the binary
system. If, however, mass transfer is thermally unstable, the
evolution of a LMXB, undergoing mass transfer cycles, is totally
different from that without irradiation. Mass transfer is spasmodic
with phases of high mass transfer driven by the thermal expansion
of the convective envelope of the irradiated donor alternating with
phases with low or no mass transfer during which the donor readjusts 
towards thermal equilibrium of the unirradiated star. Because the
thermal timescale of the convective envelope can be rather short, the
mass transfer rate during a high state can exceed the Eddington
accretion rate of a neutron star by a large factor. And this results
in the same effects as have already been discussed in Sect.~
\ref{DiscEffects}, namely in considerable loss of mass and angular
momentum from the system and low accretion efficiency of the neutron
star. This, in turn, hinders the spin-up of the neutron star and thus
possibly the formation of a ms-pulsar. In the low state the binary is
more or less detached and the mass transfer rate is either very low or
virtually zero. Thereby the duration of the low state is proportional
to the timescale on which mass transfer is driven and, therefore, can
be very long. And, of course, during the long phases with little or no
mass transfer the previously spun-up neutron star can be spun down
again by the action of the propeller mechanism (see below Sect.
\ref{propeller}). 

\begin{figure}
\begin{center}
\includegraphics*[width=8.0cm]{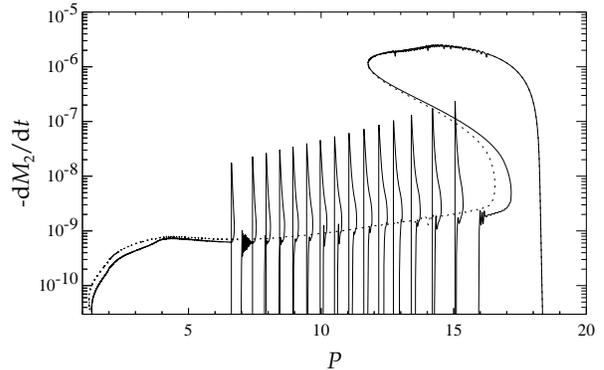}
\end{center}
\caption{Mass transfer rate (in {$\rm M_{\odot}/yr$}) as a function of
the orbital period (in $\rm d$) of a model LMXB with (full line) and
without (dotted line) taking irradiation into account. For details see
text. From \citet{B&R04}.} 
\label{MSLMXBevol}
\end{figure}

An example (from \citet{B&R04}) for how different the evolution of a
neutron star LMXB can be, depending on whether irradiation is taken
into account or not, is shown in Fig.~\ref{MSLMXBevol}. The para\-me\-ters
and assumptions used for these calculations were as follows: for the
neutron star: initial mass $M_{\rm NS} = 1.4 {\rm M_{\odot}}$, radius 
$R_{\rm NS} = 10^6$cm; for the donor star at the onset of mass
transfer $M_{\rm d} = 3 {\rm M_{\odot}}$ and central hydrogen content
$X_{\rm c} = 0.36$; angular momentum loss by gravitational radiation
and magnetic braking according to \citet{V&Z81} with $f_{\rm VZ} = 1$;
conservative mass transfer as long as $\dot M_{\rm tr} < 
\dot M_{\rm Edd} = 2~10^{-8} {\rm M_{\odot}/yr}$, and loss of mass and
angular momentum in the Jeans mode with $\dot M_{\rm loss} =
\dot M_{\rm tr} - \dot M_{\rm Edd}$  if $\dot M_{\rm tr} >
\dot M_{\rm Edd}$; irradiation of the donor by a point source at the
location of the neutron star with a flux equal to 10\% of the
perpendicular component of the isotropic flux resulting from the
accretion luminosity $L_{\rm acc} = G M_{\rm NS} {\dot M_{\rm accr}}/
R_{\rm NS}$, and ignoring the shadow cast by the disc onto the donor. 
As can be seen from Fig. \ref{MSLMXBevol}, after an initial phase of
thermally unstable mass transfer, if irradiation is taken into
account irradiation-driven mass transfer cycles start to appear.
Thereby the mass transfer rate during the high state of a mass
transfer cycle can exceed the time-averaged mass transfer rate by up
to two orders of magnitude and $\dot M_{\rm Edd}$ by up to a factor of
10. This shows that it makes really a big difference for the long-term
evolution of a LMXB whether or not irradiation-driven mass transfer
cycles do occur. 

Such differences are even more extreme if the donor star is a
giant. There are two main reasons for this: First, as can be seen from
Eq. (\ref{DR/R}), with $\rho \approx 0.5$ the amplitude of the
irradiation effect for a given value of $s_{\rm eff}$ is much higher
than for a main sequence star, where $\rho \approx 0.1$. This is a
direct consequence of the core mass-luminosity relation: upon expansion
of the convective envelope of a giant the nuclear energy generation is
not quenched, in contrast to what happens in main sequence
stars. Second, because of the much larger radius $R$ and luminosity
$L$ of a giant compared to a typical low-mass main sequence star, the
thermal timescale of the convective envelope $\tau_{\rm ce}$ (being
proportional to the Kelvin-Helmholtz time $\tau_{\rm KH} \propto
(RL)^{-1}$) is much shorter and, therefore, the irradiation-driven
mass transfer rate $\dot M_{\rm tr} \propto M/{\tau_{\rm ce}}$
correspondingly higher.   

\begin{figure}
\begin{center}
\includegraphics*[width=8.0cm]{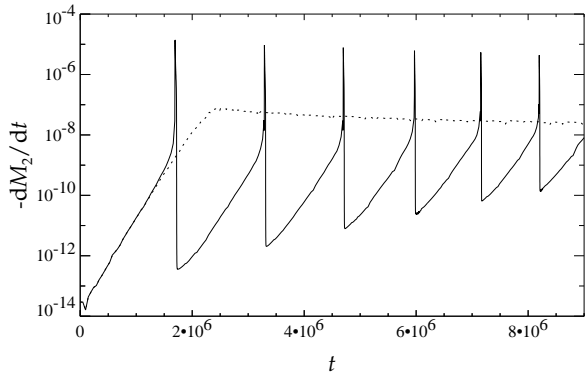}
\end{center}
\caption{Mass transfer rate (in {$\rm M_{\odot}/yr$}) as a function of
time $t$ (in yr) of a model LMXB with a giant donor with (full line)
and without (dotted line) taking irradiation into account. For details
see text. From \citet{B&R04}.}
\label{giantLMXBevol}
\end{figure}

This is illustrated by the results of a computation by \citet{B&R04}
shown in Fig.~\ref{giantLMXBevol}. In this case the parameters and
assumptions used for the model calculation were as follows: for the
neutron star: initial mass $M_{\rm NS} = 1.4 {\rm M_{\odot}}$, radius 
$R_{\rm NS} = 10^6$~cm; the donor star at the onset of mass transfer
is a giant with $M_{\rm d} = 0.8 {\rm M_{\odot}}$ and a radius
$R_{\rm d} = 25.8 {\rm R_{\odot}}$; mass transfer is driven by 
nuclear evolution of the donor and assumed to be conservative;
irradiation of the donor by a point source at the location of the
neutron star with a flux equal to 10\% of the perpendicular component 
of the isotropic flux resulting from the accretion luminosity
$L_{\rm acc}$, and ignoring the shadow cast by the disc onto the
donor.  

In this example, if mass transfer cycles do occur, mass transfer
proceeds in very short bursts during which exceedingly high mass
transfer rates are reached (up to $\dot M_{\rm tr} \approx 10^{-5}
{\rm M_{\odot}/yr}$), each of which is followed by a long detached
phase lasting typically $\sim 10^6$~yr. 

Whether or not irradiation-driven mass transfer cycles do occur also
makes a big difference when it comes to estimating the intrinsic 
number of LMXBs e.g. in the Galaxy. Since the high state of such a
mass transfer cycle lasts much longer than the ${\sim}~40$~yr since
we observe LMXBs we actually see only a small fraction of those
systems undergoing mass transfer cycles, namely those which are
currently in a high state. And that fraction is roughly the average
duty cycle $\left<d\right> = \left< t_{\rm high}/t_{\rm cycle}
\right>$. Given that the peak mass transfer rate can exceed the
time-averaged mass transfer rate by several orders of magnitude (see
Figs. \ref{MSLMXBevol} and \ref{giantLMXBevol}), the occurrence of
irradiation-driven mass transfer cycles would also imply the
existence of a large and hidden population of LMXBs being in the low
state of the cycle. 

\section{Consequences for the secular evolution of black hole LMXBs}
\label{BHLMXBs}

\subsection{Consequences from disc irradiation}
\label{BHLMXBs.1} 

When comparing Eq.~(\ref{Rhvisc}) with Eq.~(\ref{RhirrBH}) we see that
also in the case of a black hole LMXB $R_{\rm h, irr} > 
R_{\rm h, visc}$ for any value of $\dot M_{\rm tr}$ of interest and
that also in this case irradiation of the disc widens the parameter
space for systems with stable disc accretion. Following again
\citet{King&Ritter98} and using standard parameters for a black hole
LMXB, and assuming $R_{\rm d}$ to be 80\% of the Roche radius of the
compact component the critical orbital period separating transient
from non-transient systems is 
\begin{equation}
P_{\rm crit} \approx 2.3~{\rm d}~{m_{\rm c}}^{-0.675} 
                     {m_{\rm d}}^{0.175} {\dot M_{\rm tr}}^{3/4}\,.
\label{PcritBH}
\end{equation}
As in the case of neutron star LMXBs disc irradiation in black hole
LMXBs is of little consequence for the long-term evolution as long as
$P_{\rm orb} < P_{\rm crit}$, i.e. as long as disc accretion is
thermally stable, apart from the fact that stable disc accretion is a
pre\-re\-qui\-site for the occurrence of irradiation-driven mass transfer
cycles. However, as has first been pointed out by \citet{KKB96},
practically all black hole LMXBs with a non-compact donor star must be
transient. Apart from the fact that in a black hole LMXB irradiation
is less efficient than in a neutron star LMXB and, therefore,
$P_{\rm crit}$ is smaller, the main reason for this is that the mass
of the black hole $M_{\rm BH}$ is significantly higher than the mass
of a typical neutron star, i.e. $\left<M_{\rm BH}\right> \approx 3 -
10 \left<M_{\rm NS}\right>$. Therefore, the mass ratio $M_{\rm c}/
M_{\rm d}$ is larger by a corresponding factor, and this, in turn,
results in lower mass transfer rates because, on the one hand,
mass transfer is the more stable the larger $M_{\rm c}/M_{\rm d}$, 
and, on the other hand, because the rate of loss of orbital angular
momentum is smaller for larger $M_{\rm c}/M_{\rm d}$ (at least for the
often-used prescription of ``magnetic braking'' by \citet{V&Z81}). 

Another consequence of having a black hole accretor rather than a
neutron star is that the relevant Eddington accretion rate $\dot
M_{\rm Edd} \propto M_{\rm c}$ is higher, on average by a factor
$\left<M_{\rm BH}/M_{\rm NS}\right>  \approx 3 - 10 $. And
this means that supper-Eddington accretion is less likely to
occur. This applies even more to the (hypothetical) non-transient
systems. In the end this means that in black hole LMXBs the mass and
consequential angular momentum losses are systematically smaller than
in neutron star LMXBs. Yet because $R_{\rm h, irr} > R_{\rm visc}$ a
larger part of the disc is actively involved in an outburst and thus
during an outburst the mass flow rate through the disc is
systematically higher than what one would have in a non-irradiated
disc. As a consequence, super-Eddington rates and associated mass and
consequential angular momentum loss can nevertheless result in systems
where none would occur in the absence of irradiation. 

\subsection{Consequences from irradiating the donor} 
\label{BHLMXBs.2}

Because, as we have seen in the previous section, practically all
black hole LMXBs are transient, irradiation of the donor star is
intermittent on timescales short compared to the thermal timescale of
the donor's convectiec envelope $\tau_{\rm ce}$. For reasons which we
shall discuss in the next section, irradiation-driven mass transfer
cycles are not expected to occur under these circumstances. 

\section{Disc instabilities and mass transfer cycles at the same time
in the same system?}
\label{DI&IMTC}

So far I have dealt with disc instabilities and irradiation-driven
mass transfer cycles as separate issues. The question is now whether
both could occur in the same system at the same time. And, as I shall
show below, the answer is: {\bf No}. In order to understand why this
is so it is necessary to first have a closer look at the prerequisits
for irradiation-driven mass transfer cycles to occur: 

\subsection{Conditions for irradiation-driven mass transfer cycles} 
\label{SectDI&IMTC.1} 

The list given below of conditions which favour the occurrence of mass
transfer cycles follows directly from the discussion of thermal
stability of mass transfer as given e.g. in RZK, \citet{KFKR96},
\citet{KFKR97}, and \citet{B&R04}. 
\begin{enumerate}
\item {\em Sustained irradiation with a flux $F_{\rm irr} \approx 
      F_{\rm int}$ over at least a thermal timescale of the
      convective envelope $\tau_{\rm ce}$.}

      If this condition is violated, i.e. if either $F_{\rm irr}$ is
      small or irradiation is intermittent, i.e. if phases of
      irradiation with high flux but short duration $\Delta
      t_{\rm irr} < \tau_{\rm ce}$ alternate with phases with very
      little or no irradiation, blocking of intrinsic flux is 
      inefficient . The reason for this is the run of the function 
      $F_{\rm int}(F_{\rm irr})$, examples of which are shown in
      Fig. \ref{Fint(Firr)}. Because of the non-linearity of this
      function, i.e. because no more than the intrinsic flux $F_0$ of
      the unirradiated star can be blocked, the maxi\-mum amount of
      energy blocked for a given amount of accretion luminosity is
      achieved by continuous irradiation with the time-averaged flux.
      And although the blocking of intrinsic luminosity is highest for
      the highest irradiating fluxes, continuous irradiation with very
      high fluxes $F_{\rm irr} \gg F_{\rm int}$ does not help either
      because for mass transfer to be unstable $-d F_{\rm int}/
      d F_{\rm irr}$ must not be too small. This is the case only 
      for fluxes $F_{\rm irr}/{F_0} \lesssim 1,\dots, {\rm few}$. On
      the other hand, for large $F_{\rm irr},~d F_{\rm int}/
      d F_{\rm irr} \rightarrow~0$, and mass transfer is stable
      despite irradiation. 

      The extent to which intermittent irradiation can suppress the
      occurence of irradiation-driven mass transfer cycles has been
      examined by \citet{KFKR97} by adopting a simplified $F_{\rm
      int}(F_{\rm irr})$-relation and a ``top hat model'' for the
      temporal variation of the accretion rate. Not surprisingly, it
      is found that mass transfer is the more thermally stable the 
      smaller the duty cycle of the intermittency of irradiation is. 
\item {\em A small value of $\tau_{\rm ce}/t_{\rm dr}$}, where
      $t_{\rm dr}$ is the timescale on which mass transfer is driven,
      i.e. $t_{\rm dr} = (1/{t_{\rm nuc}} + 2/{t_{\rm J}})^{-1}$. Here
      $t_{\rm nuc}$ is the timescale on which the radius of the donor
      grows in response to nuclear evolution alone, and $t_{\rm J}$
      the timescale on which orbital angular momentum (in the absence
      of mass transfer) is lost. For this condition to be fulfilled
      either $\tau_{\rm ce}$ has to be sufficiently small or
      $t_{\rm dr}$ not too short. The former is the case either for
      main sequence stars having a relatively shallow convective
      envelope or for giants, the latter for binary systems losing
      orbital angular momentum at the minimum possible rate, i.e. via
      gravitational radiation only. 
\item {\em A small photospheric scale height $H$ of the donor, more
      precisely $H/R_{\rm d} \ll 1$.} 

      The smaller $H/R_{\rm d}$ the larger the derivative
      $d \dot M_{\rm tr}/d R_{\rm d}$  (e.g. \citep{R88}), i.e. the
      more violent the reaction of the system upon an
      irradiation-driven change of the donor's radius. Typical values
      are $H/R \approx 10^{-4}$ for main sequence stars and $10^{-2}
      \lesssim H/R \lesssim 10^{-3}$ for giants. 
\item {\em A small value of $R_{\rm c}/R_{\rm d}$.} 

      Since for a LMXB $R_{\rm c}$ is already minimal, $R_{\rm c}/
      R_{\rm d}$ is necessarily small, unless the donor is a compact
      star, and systematically smaller for a giant donor than for a
      main sequence star.
\item {\em A sufficiently large fraction of the stellar surface has to
      be  exposed to a flux $F_{\rm irr} \approx F_{\rm int}$.} 

      If the source of irradiation is modelled as a point source and
      the shadow cast by the disc onto the donor star is ignored, a
      fraction $s_{\rm PS} = R_{\rm d}/a \approx 0.3 - 0.4$ of the
      donor's surface ``sees'' the source. Here $a$ is the orbital
      separation of the two stars. However, the effective fraction of
      the surface  $s_{\rm eff}$ over which $F_{\rm int}$ is blocked
      can be considerably smaller than $s_{\rm PS}$ because the
      perpendicular component of the irradiating flux (this is the
      quantity that is relevant here) varies considerably from the
      substellar point to the termiator: If $F_{\rm irr} \approx
      F_{\rm int}$ near the substellar point, then $F_{\rm irr}$ is
      small (and so is the blocking of intrinsic flux) far from it. On
      the other hand, if $F_{\rm irr} \approx F_{\rm int}$ far from
      the substellar point then closer to it $F_{\rm irr} \gg
      F_{\rm int}$, i.e. the blocking effect saturates, and
      $s_{\rm eff} \rightarrow s_{\rm PS}$. Things get even more
      unfavourable if the disc shadow is taken into account: The fact
      that few LMXBs show deep eclipses has always been interpreted as
      evidence for the outer parts of the disc to be flared to the
      extent that practically no part of the donor ``sees'' the disc's
      center. Turning the argument around this means that in a LMXB no
      part or at most small regions near the polar caps can be
      directly  rradiated by a point source located at the disc's
      center.  
\end{enumerate}

\subsection{Consequences for systems with a giant donor} 
\label{DI&IMTC.2}
As we have seen in Sect. \ref{NSLMXBs.1} in systems with a giant donor
fulfilling the criterion $P_{\rm orb} > P_{\rm crit}$, where
$P_{\rm crit}$ is given by either Eq.~(\ref{PcritNS}) or
Eq.~(\ref{PcritBH}), disc instabilities are unavoidable. Hence the
irradiation resulting from accretion is intermittent on a timescale
which is essentially given by the low state viscous timescale of the
``active'' accretion disc \citep{Ritter&King01}. And that, in turn, is
of the order of $\sim 10 - 10^3$~yr, and thus much shorter than
$\tau_{\rm ce}$ of the associated giant donor. Therefore, systems with
a giant donor and $P_{\rm orb} > P_{\rm crit}$ violate the first of
the  above criteria, and this is already enough to suppress
irradiation-driven mass transfer cycles in LMXBs with $P_{\rm orb} >
P_{\rm crit}$. 

In this sense, the calculation shown as a full line in
Fig.~\ref{giantLMXBevol} is irrelevant. After what has been said
above, systems undergoing such an evolution should not exist. And we
have indirect evidence that this is indeed the case, namely the
existence of ms-pulsars in binary systems with orbital periods even as
long as a few $10^2$~d. If such LMXBs would really evolve with
irradiation-driven mass transfer cycles and behave as shown in 
Fig.~\ref{giantLMXBevol} (full line), the total amount of mass the
neutron star could accrete during the comparatively very short high
states would be very small and the angular momentum accreted with it
insufficient to spin the neutron star up to a spin period of a few
milliseconds.  

\subsection{Consequences for systems with a main sequence donor} 
\label{DI&IMTC.3} 
As we have seen, virtually all black hole LMXBs, even those with a 
main sequence donor, must be transient (c.f. Eq.~(\ref{PcritBH})).
Therefore, irradiation is intermittent and irradiation-driven mass
transfer cycles are suppressed. 

On the other hand, neutron star LMXBs with stable disc accretion are
possible (c.f. Eq.~(\ref{PcritNS})). In fact, irradiation, by
suppressing disc instabilities, widens the parameter space for stable
disc accretion, and for such systems the occurrence of
irradiation-driven mass transfer cycles cannot be ruled out a priori.
As we shall discuss in the next section, the main problem here is
calculating $s_{\rm eff}$, i.e. $F_{\rm irr}$, in the context of a
reliable model of a LMXB. 

Finally, also in transient neutron star LMXBs irradiation-driven mass
transfer cycles are suppressed because irradiation of the donor is
intermittent.  

\section{Open problems}
\label{open.problems}
\subsection{Direct versus indirect irradiation}
\label{irrad.mode}
One of the most serious problems, if not the most serious one, arising
in the contex of our topic is how to calculate $F_{\rm irr}$ for each
surface element of the object of interest, i.e. either the (outer
parts of the) accretion disc or the donor star. As has e.g. been shown
by \citet{DHL01} this is already a non-trivial task in one of the
simplest cases imaginable, i.e. a planar, axisymmetric concave disc
being irradiated by a central point source. 

The situation is considerably more complicated when considering that
in going to outburst an initially cool disc is transformed into a hot
disc out to some radius $R_{\rm h, irr} > R_{\rm h, visc}$ (cf. Eqs.
(\ref{Rhvisc})  and, respectively, (\ref{RhirrNS}) or (\ref{RhirrBH})).
Whereas a central point source can keep the disc hot, i.e. at
$T_{\rm eff} > T_{\rm H}$, in the region $R_{\rm h, irr} > r >
R_{\rm h, visc}$, it cannot transform the same region from the cool
state into the hot state because in the cool state these regions are
in the (point source) shadow cast by the inner, hot parts of the
disc. Since the observed long durations of the outbursts of X--ray
transients are a natural outcome of assuming that in outburst the disc
is in the hot state out to some radius $R_{\rm h, irr}  >
R_{\rm h, visc}$ (see e.g. \citet{King&Ritter98} and \citet{DHL01}),
we are practically forced to the conclusion that indirect irradiation,
i.e. scattered light, must be involved. And, as the following estimate
shows, this is not inconceivable because only a small fraction of the
isotropic accretion flux $F_{\rm accr} = L_{\rm accr}/4 \pi r^2$ needs
to be scattered towards the disc in order to raise its temperature
above $T_{\rm H}$ out to some radius $r \leqq R_{\rm h, irr}$. Since
the required scattered flux is $F_{\rm irr} \approx \sigma
{T_{\rm H}}^4$ we find that, assuming an X-ray albedo of $\sim 0.1$,
$F_{\rm irr}/F_{\rm accr} \lesssim 10^{-3} (r/R_{\odot})^2$ for a
neutron star or a black hole LMXB. (Note that for this estimate using
Eq. (\ref{RhirrNS}) or (\ref{RhirrBH}) for $R_{\rm h, irr}$ is
inappropriate because these expressions have been derived by assuming
direct irradiation.) Invoking indirect irradiation also requires that
the irradiating source, i.e. the scattering corona, is sufficiently
extended. Of course, once the disc has been brought to the hot state
in which it is concave for radii $r < R_{\rm h, irr}$ (whereby it is
unclear whether in this case $R_{\rm h, irr}$ is adequately
approximated by respectively Eq.(\ref{RhirrNS}) or (\ref{RhirrBH})),
direct irradiation will also contribute, in addition to scattered
light.   

Calculating $F_{\rm irr}$ is not only a problem when dealing with
irradiated discs in X-ray transients, but even more so when dealing
with an irradiated donor star in the context of irradiation-driven
mass transfer cycles. First, it is important to note that for bright
LMXBs direct irradiation probably does not work. On the one hand, in
the presence of a stationary hot accretion disc which casts a broad
point source shadow onto the donor star, only relatively small areas
near the poles of the donor's facing hemisphere are directly
irradiated by the central source, and that at near grazing incidence.
Although details have not been worked out so far, it is very probable
that this is insufficient to destabilize mass transfer. On the other
hand, even if one ignores the shadow cast by the disc onto the donor
star mass transfer is not likely to be unstable. The reason
for this has already been discussed above (point (v) in Sect.
\ref{SectDI&IMTC.1}): in LMXBs with a low-mass main sequence donor
\footnote{as we have argued above, only for such systems
irradiation-driven mass transfer cycles could possibly occur}, i.e.
with $M_{\rm d} \lesssim 1~M_{\odot}$, the ratio $F_{\rm irr}/
F_{\rm int}$ near the substellar point is typically much larger than
unity even for small values of the X-ray albedo (of order $\lesssim
0.1$). And this is also the case for most parts of the facing
hemisphere except for small regions at high latitude where irradiation
is nearly grazing. Therefore, as far as the stability of mass transfer
is concerned, the situation is not unlike the one  where the disc's
shadow is taken into account: only a small fraction of the stellar
surface is exposed to irradiation and at the same time sufficiently
sensitive to {\em changes} in $F_{\rm irr}$. 

Since direct irradiation is unlikely to destabilize mass transfer in a
LMXB we are now going to examine whether indirect irradiation could do
the job. The first thing to note in this context is that if in wide,
transient LMXBs scattered X-rays are intense enough to raise the
effective temperature of the outer parts of an otherwise cool disc
above $T_{\rm H} \approx 6500$~K they will also be intense enough to 
result in ratios $F_{\rm irr}/F_{\rm int} \gtrsim 1$ on the donor
star of a typi\-cal short-period, non-transient LMXB. As we had argued
above (point (i) in Sect. \ref{SectDI&IMTC.1}) such values of 
$F_{\rm irr}/ F_{\rm int}$ are optimal for inducing mass transfer
cycles. The other question is whether a sufficiently large fraction of
the donor's surface is exposed to scattered light. This depends
entirely on the size of the scattering corona. If it is small compared
to the orbital separation, the disc will nevertheless cast an extended
shadow onto the donor, and irradiation, affecting too small an area,
will probably not destabilize mass transfer. If, on the other hand,
the irradiating source is sufficiently extended, not only will the
shadowing of the disc be much less significant. In addition, regions
on the donor star which are well beyond the point source terminator
could be sig\-ni\-fi\-cantly affected by indirect irradiation. Clearly,
calculating $F_{\rm irr}$ over the surface of the donor star under
these circumstances is no simple task. And, not surprisingly, hitherto
no  such calculations, though urgently needed, have been carried
out. For this reason it is currently also impossible to say whether
LMXBs could undergo irradiation-driven mass transfer cycles.

\subsection{The disc instability model}
Adequate modelling of the spin-up of a neutron star to a ms-pulsar in
a long-period, transient LMXB requires detailed knowledge of how (how
much and for how long) mass is flowing through the accretion disc onto
the neutron star during an ouburst. This necessarily involves the 
thermal-viscous disc instability model for discs subject to
irradiation during an outburst. And, although numerical simulations of
one or a few such outbursts for a few sets of parameters have been
carried out (see e.g. \citet{DHL01}, and references therein), such 
calculations are not sufficient for the task at hand: after all we are
talking here about the entire phase of mass transfer from a giant
donor which can last up to $\sim 10^8$~yr (e.g. \citet{Ritter99}), 
i.e. a time during which the disc undergoes a huge number of outbursts
under secularly changing conditions. No question that, at least at
present, this could be dealt with by means of a sequence of detailed
numerical disc instability model calculations . What is really needed
is a reasonably simple yet sufficiently accurate analytical or
semi-analytical model of disc instabilities which provides the entire 
manifold of solutions. Whereas a viable model for the outburst phase,
during which the active disc (that is the  part of the disc involved
in an outburst) is in a quasi-stationary state, can be formulated
\citep{Ritter&King01}, this has so far not been possible for the
quiescent phase during which the active disc is not nearly
stationary. On top of that any viable model has also to take into
account that all the important properties of an outburst cycle are
strongly influenced by the fact that during quiescence matter in the
central part of the disc evaporates into an advection-dominated
accretion flow (\citet{MMH94}, \citet{LMMH97}, \citet{MLMH00},
\citet{DHL01}, and references therein). Thereby a central hole is
formed, the size of which essentially determines the storage capacity
of the disc and thus the duration of quiescence. Until such a model
becomes available, the spin-up of a neutron star to a ms-pulsar in a
long-period LMXB cannot be adequately modelled.

\subsection{Spinning down the neutron star}
\label{propeller}
In the contex of the formation of ms-pulsars in long-period, transient
LMXBs we should keep in mind that during the long-lasting quiescence
phases the neutron star could also be spun down by the propeller
effect \citep{Ill&Sun75}. A classical propeller, i.e. spin-down
of the neutron star, results when during an accretion phase the mass
flow rate through the disc drops so much that the magnetospheric
radius $R_{\rm M}$ exceeds the corotation radius $R_{\rm co}$. Here
the situation is different: In an outburst during which the neutron
star is spun up, the inner radius of the disc is $R_{\rm i} \approx
R_{\rm M, outb} \lesssim R_{\rm co}$, where $R_{\rm M, outb}$ is the
magnetospheric radius for a (ststionary) disc in ouburst. Once the
disc has gone in quiescence, its central parts, out to a radius
$R_{\rm ev} \sim 10^{9.5} {\rm cm}$, evaporate into an
advection-dominated coronal accretion flow \citep{MMH94}, and the
inner disc radius $R_{\rm i}$ is set by $R_{\rm ev}$ which is
typi\-cally much larger than the magnetospheric radius in quiescence
$R_{\rm M, qsc}$ calculated for the pressure of the coronal gas.
Because for typical values of the parameters of the problem
$R_{\rm M, qsc} \gg R_{\rm M, outb}$ the neutron star will be spun
down during quiescence if it is spinning not too far below the
equilibrium spin frequency corresponding to the outburst accretion
rate. To what extent this spin-down is significant remains yet to be
determined. Should it be significant then, depending on the initial
conditions of a binary, this could even prevent the formation of a
ms-pulsar.

Thus, spin-up of a neutron star to ms spin periods can occur if, on
the one hand, the spin-down during the quiescence phases is not too 
large, and, on the other hand, either the neutron star magnetic moment
is small (of order $10^{26} {\rm G~cm^3}$) from the beginning, or
decreases as a result of accretion, and the accretion efficiency is
not too small. With ongoing spin-up the radius of the 
light cylinder $R_{\rm LC} = 4.8~10^9 {\rm cm} P_{\rm spin}(s)$
becomes eventually smaller than the inner radius of the disc in
quiescence, i.e. $R_{\rm LC} < R_{\rm ev}$, and the corona becomes
exposed to the pressure of the pulsar wind. With decreasing spin
period the power of the pulsar wind $\propto \mu^2
{P_{\rm spin}}^{-4}$ grows, and one may ask whether at some point the
pulsar wind is strong enough to blow away the evaporating coronal gas.
A rough estimate using Eq. (17) of \citet{MMH94} for the pressure of
the coronal gas shows that this is indeed possible when the spin
period drops below $\sim 0.1 {\rm s}$.  Beyond that point the pulsar
will no longer be spun down by the propeller effect during quiescence,
and the spin-up process continues with higher efficiency. 

From what has just been said it should have become clear that an
adequate modelling of the spin history of a neutron star in a
transient LMXB is a very demanding task which requires a proper
treatment of the above-mentioned open problems.

\section{Conclusions}

In the foregoing sections we have discussed at some length the
question whether irradiation is important for the secular evolution of
LMXBs. Whereas there is clear observational evidence that irradiation
does change the optical appearence of LMXBs \citep{van_Paradijs94} and
the properties of the outbursts in transient systems (e.g.
\citet{King&Ritter98}, \citet{DHL01}), at present it is much less clear
to what extent irradiation of either the outer parts of the accretion
disc or the donor star influences the long-term evolution of
LMXBs. The main reason for this is that a number of important problems
which we have discussed in Sect. \ref{open.problems} need first to be
solved. 

The main effects which irradiation in a LMXB could have on its secular
evolution are: 1) enhanced loss of mass and angular momentum from the
system as a consequence of super-Eddington mass flow rates during the
outbursts of transient LMXBs. This effect is more important for
neutron star LMXBs than for black hole LMXBs. 2) For neutron star LMXBs
the higher mass loss rates in an outburst mean that a lower fraction
of the transferred mass is available for accretion onto the neutron
star and thus for spinning it up. In this way irradiation makes it
more difficult or even impossible for the neutron star to become a
ms-pulsar. 3) Irradiation of the donor star can destabilize mass
transfer and force the system to undergo irradiation-driven mass
transfer cycles, i.e. an evolution which differs drastically from that
expected without taking into account irradiation. 

We have also seen that irradiation-driven mass transfer cycles could
only occur in systems in which irradiation is sustained for a
sufficiently long time, i.e. at least of order of the thermal
timescale of the donor's convective envelope. Therefore, the
occurrence of irradiation-driven mass transfer cycles is restricted to
systems in which disc accretion is stable. 

Finally we have seen that indirect irradiation by scattered 
accretion luminosity is probably needed (and also available) for
irradiation-driven mass transfer cycles to work and, in transient
LMXBs, for transforming an initially cool disc into a hot disc beyond
the radius $R_{\rm h, visc}$ (cf. Eq. (\ref{Rhvisc})).

\section{Acknowledgements}

I am grateful to Marek Abramowicz for having invited me to, and to
Marek and his coworkers for organizing the memorable birthday 
conference  for Jean-Pierre Lasota. I am also grateful to
Dr. Friedrich Meyer for many stimulating discussions. 

% The Appendices part is started with the command \appendix;
% appendix sections are then done as normal sections
% \appendix

% \section{}
% \label{}

% Bibliographic references with the natbib package:
% Parenthetical: \citep{Bai92} produces (Bailyn 1992).
% Textual: \citet{Bai95} produces Bailyn et al. (1995).
% An affix and part of a reference:
%   \citep[e.g.][Ch. 2]{Bar76}
%   produces (e.g. Barnes et al. 1976, Ch. 2).


\begin{thebibliography}{}

% \bibitem[Names(Year)]{label} or \bibitem[Names(Year)Long names]{label}.
% (\harvarditem{Name}{Year}{label} is also supported.)
% Text of bibliographic item

\bibitem[B\"uning \& Ritter(2004)]{B&R04}
        B\"uning, A, \& Ritter, H., 2004, A\&A 423, 281.
\bibitem[Dubus, Hameury \& Lasota(2001)]{DHL01}
        Dubus, G., Hameury, J.-M., Lasota, J.-P., 2001, A\&A 373, 251. 
\bibitem[Hameury \& Ritter(1997)]{Hameury&Ritter97}
        Hameury, J.-M., Ritter, H., 1997, A\&AS 123, 273. 
\bibitem[Illarionov \& Sunyaev(1975)]{Ill&Sun75}
        Illarionov, A.F., \& Sunyaev, R.A., 1975, A\&A 39, 185. 
\bibitem[King et al.(1996)]{KFKR96}
        King, A.R., Frank, J., Kolb, U., Ritter, H., 1996, ApJ 467,
        761. 
\bibitem[King, Kolb \& Burderi(1996)]{KKB96} 
        King, A.R., Kolb, U., Burderi, L. 1996, ApJ 464, L127.
\bibitem[King et al.(1997)]{KFKR97}
        King, A.R., Frank, J., Kolb, U., Ritter, H., 1996, ApJ 482,
        919. 
\bibitem[King \& Ritter(1998)]{King&Ritter98}
        King, A.R., \& Ritter, H., 1998, MNRAS 293, L42.  
\bibitem[Lasota(2001)]{Lasota01}
        Lasota, J.-P., 2001, NAR 45, 449. 
\bibitem[Liu, Meyer \& Meyer-Hofmeister(1997)]{LMMH97}
        Liu, B.F., Meyer, F., \& Meyer-Hofmeister, E., 1997, A\&A 328,
        247. 
\bibitem[Meyer, Liu \& Meyer-Hofmeister(2000)]{MLMH00} 
        Meyer, F., Liu, B.F., \& Meyer-Hofmeister, E., 2000, A\&A 361,
        175.
\bibitem[Meyer \& Meyer-Hofmeister(1994)]{MMH94}
        Meyer, F. \& Meyer-Hofmeister, E., 1994, A\&A 288, 175. 
\bibitem[Podsiadlowski(1991)]{Podsiadlowski91}
        Podsiadlowski, Ph., 1991, Nature 350, 136. 
\bibitem[Ritter(1988)]{R88}
        Ritter, H., 1988, A\&A 202, 93.
\bibitem[Ritter(1999)]{Ritter99}
        Ritter, H., 1999, MNRAS 309, 360.
\bibitem[Ritter, Zhang \& Kolb (2000)]{RZK00}
        Ritter, H., Zhang, Z.-Y., \& Kolb, U., 2000, A\&A 360, 969. 
\bibitem [Ritter \& King(2001)]{Ritter&King01}
        Ritter, H., \& King, A.R., 2001, in: Evolution of Binary and
        Multiple Star Systems, eds. Ph. Podsiadlowski, S. Rappaport,
        A.R. King, F. D'Antona, \& L. Burderi, ASP Conf. Ser.,
        Vol. 229, 423.
\bibitem[van Paradijs(1994)]{van_Paradijs94}
        van Paradijs, J., 1996, ApJ 464, L139.  
\bibitem[Verbunt \& Zwaan(1981)]{V&Z81}
        Verbunt, F., \& Zwaan C., 1981, A\&A 100, L7. 
\end{thebibliography}
\end{document}